\documentclass[showpacs, preprintnumbers, nofootinbib, aps, prd, superscriptaddress,10pt, showkeys, notitlepage, twocolumn]{revtex4-1}

\usepackage{graphicx,amssymb,amsmath,amsthm,amsfonts,epsfig}

\usepackage[linktocpage,breaklinks]{hyperref}
\usepackage[usenames,dvipsnames]{color}
\usepackage{epstopdf}
\usepackage{aas_macros}
\usepackage{pifont}
\definecolor{darkred}{rgb}{0.5,0,0}
\definecolor{darkgreen}{rgb}{0,0.5,0}
\definecolor{darkblue}{rgb}{0,0,0.5}
\definecolor{prussian}{rgb}{0.0, 0.19, 0.33}
\definecolor{richelectricblue}{rgb}{0.03, 0.57, 0.82}
\definecolor{teal}{rgb}{0.0, 0.5, 0.5}
\definecolor{mediumseagreen}{rgb}{0.24, 0.7, 0.44}
\definecolor{lust}{rgb}{0.9, 0.13, 0.13}
\definecolor{ballblue}{rgb}{0.13, 0.67, 0.8}
\definecolor{darkcyan}{rgb}{0.0, 0.55, 0.55}
\definecolor{mountainmeadow}{rgb}{0.19, 0.73, 0.56}
\definecolor{palecarmine}{rgb}{0.69, 0.25, 0.21}
\definecolor{richcarmine}{rgb}{0.84, 0.0, 0.25}
\definecolor{tangelo}{rgb}{0.98, 0.3, 0.0}
\definecolor{venetian}{rgb}{0.784,0.031,0.082}
\definecolor{bdfrance}{rgb}{0.192,0.549,0.906}

\hypersetup{colorlinks=true, citecolor=venetian,
linkcolor=bdfrance, urlcolor=lust}
\usepackage{amsmath,amssymb}
\usepackage{tensor}
\usepackage{mathtools}
\usepackage{amsbsy}
\usepackage{bm}
\usepackage{float}

\newcommand{\be}{\begin{equation}}
\newcommand{\ee}{\end{equation}}
\newcommand{\bea}{\begin{eqnarray}}
\newcommand{\eea}{\end{eqnarray}}
\newcommand{\nn}{\nonumber}

\newcommand{\p}{\prime}
\newcommand{\pp}{\prime\prime}

\newcommand{\cC}{{\cal C}}

\newcommand{\cD}{{\cal D}}

\begin{document}

\title{On the connection of spacetime separability and spherical photon orbits}

\begin{abstract}

The notion of non-equatorial spherical photon orbits is among the very special properties of the Kerr spacetime 
of rotating black holes and is one that leaves a clear mark on the electromagnetic and gravitational wave
signature of these objects. In principle,  one could use observations like the shadow a black hole
casts when is externally illuminated or the gravitational quasi-normal mode ringdown of merging black holes 
in order to identify the Kerr metric via its light-ring and spherical photon orbit structure, or perhaps look for the presence
of more exotic non-Kerr ultracompact objects. This approach would require some understanding of 
how circular photon orbits behave in alternative (and far less special) non-Kerr spacetimes. 
In this letter we explore the connection between the existence of spherical photon orbits and the separability of 
a general stationary and axisymmetric spacetime. We show that a spacetime cannot be separable if it possesses 
an equatorial photon ring but at the same time does not admit (in any coordinate system) non-equatorial spherical photon orbits. 
As a result, the separability-circularity connection could serve as a non-Kerr diagnostic of black hole candidates. 
\end{abstract}

\author{George Pappas}
\email{georgios.pappas@roma1.infn.it}
\affiliation{Dipartimento di Fisica,  ``Sapienza'' Universit\'a di Roma \& Sezione INFN Roma1, 
Piazzale Aldo Moro 5, 00185, Roma, Italy}

\author{Kostas Glampedakis}
\email{kostas@um.es}
\affiliation{Departamento de F\'isica, Universidad de Murcia, Murcia, E-30100, Spain}
\affiliation{Theoretical Astrophysics, University of T\"ubingen, Auf der Morgenstelle 10, T\"ubingen, D-72076, Germany}
 
\date{\today}
 
\maketitle



\emph{Context.  ---} 
Gravitational wave (GW) astronomy has finally become a tangible reality after the first direct detection of
signals from merging black holes by the Advanced LIGO/VIRGO  collaboration~\cite{Abbott:2016blz, Abbott:2016nmj, 
Abbott:2017vtc,GW170814PhysRevLett}. 
This new observational window holds promise of unprecedented tests of general relativity (GR) 
in strong gravity environments~\cite{Yunes:2013dva,Berti:2015itd,TheLIGOScientific:2016src}. 

A tantalising prospect of future GW observations is the  `beyond reasonable doubt'  identification
of the merging compact objects as Kerr black holes, as so uniquely predicted by Einstein's theory. 
What we presently call `black hole candidates' are most likely Kerr black holes, but observations still cannot rule 
out alternatives such as black holes of other -- rivals to GR -- theories of gravity, or ultracompact objects (UCOs) 
made of some kind of exotic matter. 

Tests of the so-called Kerr hypothesis with merging black hole binaries could be based on either the inspiral signal 
\cite{Berti2018GRGa} or the post-merger quasi-normal mode (QNM) ringdown \cite{Berti2018GRGb}, or a combination 
of these two. The technique of `black hole spectroscopy' rests on the measurement of a sufficient number of 
QNM frequencies and damping rates in the ringdown signal \cite{Detweiler:1980gk,Dreyer:2003bv,Berti:2005ys}
as an unequivocal indicator of the Kerr spacetime, and is facilitated by a detailed understanding of  
black hole QNM spectra (see~\cite{Kokkotas:1999bd,Berti:2009kk} for a review). 

Much of our intuition about the ringdown of Kerr black holes is based on the eikonal limit approximation: this relates 
the ringdown frequency and damping rate to the properties of the unstable photon circular orbit (usually dubbed the `photon ring')
with a surprisingly good numerical accuracy~\cite{Goebel:1972,Ferrari:1984zz, Kokkotas:1999bd,Berti:2009kk,Cardoso:2008bp,
Dolan:2010wr, Yang:2012he,Yang:2012pj,Yang:2013uba}. The basic mental picture is that of wavepackets temporarily trapped in orbit 
in the vicinity of the photon ring, slowly leaking towards infinity and the event horizon. 
Recent work~\cite{Cardoso:2016rao, Cardoso:2016oxy, Mark:2017zmc,Glampedakis2018PhRvD} has extended the 
photon ring/ringdown link to the case of black hole-mimicking UCOs as a way of approximating their largely unknown
QNM spectra. 

In the entirely different context of photon astronomy and supermassive black holes in galactic nuclei, 
photon circular orbits come to play a key role in the formation of the `shadow' that these systems cast  
when they are illuminated by e.g. a hot accretion disk. Resolving the shape and boundary of the shadow of our galactic 
Sgr $\mbox{A}^*$ black hole is one of the main objectives of the Event Horizon Telescope (EHT) \cite{Broderick2014ApJ}. 
This kind of observations could be used to probe possible deviations from the Kerr metric~\cite{Johannsen2010ApJ,Johannsen2013ApJ,
Cunha2015PhRvL,Cunha2016IJMPD, Cunha2016PhRvD, Cunha2017PhysRevD,Cunha2017PhysRevLett}.

Motivated by the importance of photon circular orbits in black hole physics, in this letter we investigate to what extent 
these orbits exist in an arbitrary  axisymmetric-stationary spacetime when motion is not confined in the equatorial plane. 
Based on the formulation of necessary conditions for the existence of a generalised class of non-equatorial circular
photon orbits, our analysis reveals an intriguing  (and, we believe, so far unnoticed) connection between \emph{spherical} 
photon orbits and the \emph{separability} of a spacetime. This connection implies that a spacetime \emph{cannot be separable} 
if it possesses an equatorial light-ring but does not admit spherical photon orbits (in any coordinate system).


\emph{Formalism for null geodesics.  ---}
We consider an arbitrary axisymmetric, stationary and equatorially symmetric 
spacetime described by a metric $g_{\mu\nu} (r,\theta)$ in a spherical-like coordinate system, 
and we further assume that the spacetime is circular, i.e., that the 2-dimensional surfaces orthogonal 
to the Killing fields are integrable \cite{wald1984}. The line element then can be written as
\be
ds^2 = g_{tt} dt^2 + g_{rr} dr^2 + 2g_{t\varphi} dt d\varphi + g_{\theta\theta} d\theta^2 + g_{\varphi\varphi} d\varphi^2.
\label{metric}
\ee
Geodesics in this spacetime have a conserved energy $E= - u_t$ and angular momentum $L = u_\varphi$ 
(along the symmetry axis). For null geodesics in particular, we can work with the impact parameter $b = L/E$ and effectively set  $E\to1$
after a rescaling  $\lambda \to E \lambda$ of the affine parameter. The norm  $u^\mu u_\mu = 0$ becomes
an equation with an effective potential (here $\cD = g_{t\varphi}^2  - g_{tt} g_{\varphi\varphi}$)
\be
 g^{rr} u_r^2 + g^{\theta\theta} u_\theta^2  = \frac{1}{\cD} \left (\, g_{tt} b^2 + 2 g_{t\varphi} b + g_{\varphi\varphi} \, \right )
\equiv   V_{\rm eff} (r,\theta,b).
\label{norm_gen}
\ee
The quadratic form of this equation implies that $V_{\rm eff}=0$ marks the zero-velocity separatrix between 
allowed and forbidden regions for geodesic motion. 

Unlike geodesic motion in Kerr where the additional existence of a third integral (Carter constant) 
results in decoupled first-order radial and meridional equations, in the general metric (\ref{metric})
one is obliged to work with the second-order geodesic equation, 
\be
\alpha_\kappa \equiv \frac{d u_\kappa}{d\lambda} = \frac{1}{2} g_{\mu\nu,\kappa} u^\mu u^\nu.
\label{geod}
\ee
For the following analysis we need the $\theta$-component of the acceleration,
\be
\alpha_\theta = \frac{1}{2}  \left (\, \frac{g_{rr,\theta}}{g_{rr}^2} u^2_r + \frac{g_{\theta\theta,\theta}}{g_{\theta\theta}^2} u_\theta^2 \, \right )
+ \frac{1}{2 } V_{{\rm eff },\theta},
\label{ath}
\ee
together with the $\lambda$-derivative of (\ref{norm_gen}), 
\begin{align}
& \frac{u_r}{g_{rr}} \left ( \,  2 \alpha_r  - \frac{g_{rr,r}}{g_{rr}^2} u_r^2 -  \frac{g_{rr,\theta}}{g_{rr} g_{\theta\theta}} u_r u_\theta \, \right ) 
\nn \\
&+ \frac{u_\theta}{g_{\theta\theta}} \left (\,   2 \alpha_\theta  - \frac{g_{\theta\theta,\theta}}{g_{\theta\theta}^2} u_\theta^2
 -  \frac{g_{\theta\theta,r}}{g_{rr} g_{\theta\theta}} u_r u_\theta   \, \right ) 
 \nn \\
& =   \frac{u_r}{g_{rr}} V_{{\rm eff},r}  + \frac{u_\theta}{g_{\theta\theta}} V_{{\rm eff},\theta}.
\label{dnorm_gen}
\end{align}
These general equations can now be applied to the particular case of circular photon orbits.


\emph{Spherical and `spheroidal' orbits.  ---}
The most familiar example of non-equatorial circular orbits are the spherical orbits of the Kerr metric
(in Boyer-Lindquist coordinates). As the name suggests, these orbits are confined on a spherical surface of radius $r_0$. 
However, one can think of a more general class of non-equatorial orbits  where motion is confined on a 
spheroidal-shaped shell $r_0 = r_0 (\theta)$ which is reflection-symmetric with respect to the equator. 
We shall call this more general type of 
orbit \emph{spheroidal}. The actual shape of the shell need not be 
spheroidal; indeed, if $r_0 (\theta)$ is not a single-valued function, the shell may be a torus-like surface in three dimensions 
(see e.g. \cite{Cunha2017PhysRevD}).

For these spheroidal orbits, the velocity components $u^r, u^\theta$ are related as 
\be
u^r = r_0^\prime u^\theta \quad \Rightarrow \quad  u_r =  \frac{g_{rr}}{g_{\theta\theta}} r_0^\prime u_\theta,
\label{uruth_circ}
\ee
where a prime stands for a derivative with respect to the argument. 
Hereafter, all functions of $r$ are to be evaluated at $r=r_0 (\theta)$. 

For the spheroidal orbit's radial acceleration we similarly obtain
\begin{align}
\alpha_r &= \frac{g_{rr}}{g_{\theta\theta}} r_0^\p \alpha_\theta + \frac{u_\theta^2}{g^3_{\theta\theta}} 
\Bigg [\, g_{\theta\theta} g_{rr} r_0^{\pp} + r_0^\p \left (\, g_{\theta\theta} g_{rr,\theta} - g_{rr} g_{\theta\theta,\theta}  \, \right ) \nn\\
&+ (r_0^\p )^2 \left (\,  g_{\theta\theta} g_{rr,r} - g_{rr} g_{\theta\theta,r}   \, \right ) \,\Bigg ].
\label{ar_circ}
\end{align}
Meanwhile, the previous Eqs. (\ref{norm_gen}), (\ref{ath}), (\ref{dnorm_gen}) become,
\begin{align}
&  \left [ g_{rr} (r^\p_0)^2 + g_{\theta\theta} \right ] u_\theta^2 = g^2_{\theta\theta} V_{\rm eff}, 
\label{norm_circ1}
\\
\nn \\
&\alpha_\theta = \frac{ u_\theta^2}{2 g_{\theta \theta}^2}  \left  [ \, g_{rr,\theta} (r_0^\p )^2  +  g_{\theta \theta,\theta} \, \right ]
+ \frac{1}{2 } V_{{\rm eff },\theta},
\label{ath_circ}
\\
\nn \\
& r_0^\p \left [ \, 2\alpha_r - \frac{r_0^\p u_\theta^2}{g_{\theta\theta}^2} \left ( \,  g_{rr,\theta} + g_{rr,r} r_0^\p  \, \right )   \, \right ] 
+ 2\alpha_\theta \nn\\
&-\frac{u^2_\theta}{g_{\theta\theta}^2} 
\left (\,  g_{\theta\theta,\theta} + g_{\theta\theta,r} r_0^\p \, \right )  =   r^\p_0 V_{\mathrm{eff},r} + V_{\mathrm{eff},\theta}.
\label{dnorm_circ1}
\end{align}
In the limit of \emph{equatorial} motion, $u_\theta = \alpha_\theta = 0$, these equations reduce to the well-known circular orbit 
conditions $V_{\rm eff} = V_{{\rm eff},r} =0$. 

We can now proceed to the derivation of a necessary `circularity' condition for spherical/spheroidal photon orbits. 
This originates from (\ref{dnorm_circ1})  after eliminating $\alpha_r$, $\alpha_\theta$ and $u^2_\theta$ with the help of 
Eqs. (\ref{ar_circ}), (\ref{ath_circ}) and (\ref{norm_circ1}), respectively. After some algebra we obtain,
\begin{align}
&0=(r_0^\p)^2 \Bigg [\,    \left (\, g_{\theta\theta} g_{rr,r} -2g_{rr} g_{\theta\theta,r} \,\right ) V_{\rm eff}  - g_{rr} g_{\theta\theta}  V_{{\rm eff},r}  \, \Bigg ]
\nn \\
& +  r_0^\p \Bigg [\,   \left (\,  2 g_{\theta\theta} g_{rr,\theta} -g_{rr} g_{\theta\theta,\theta} \, \right )  V_{\rm eff} 
+  g_{rr} g_{\theta\theta} V_{{\rm eff},\theta} \, \Bigg ] \nn\\
&+g_{rr} (r_0^\p)^3 \left ( g_{rr}  V_{\rm eff} \right )_{,\theta} + g_{\theta\theta}  \Bigg [\,    2 g_{rr} V_{\rm eff} r_0^{\pp} 
- \left ( g_{\theta\theta} V_{\rm eff} \right )_{,r}  \, \Bigg ] .
\label{circularity}
\end{align}
It is easy to verify that for spherical orbits in the Kerr metric this condition reduces to the known Kerr expression. 
It should also be noted that (\ref{circularity}) offers no information about the stability of the orbit --  this would require extra 
input from the second derivatives of $V_{\rm eff}$.


\emph{Spherical orbits and separability.  ---} 
From the condition  (\ref{circularity}) and for a metric of the form (\ref{metric}) we can establish the following 
remarkable result: if a spacetime is separable in a given coordinate system, then spherical photon orbits can 
exist, and in fact \emph{must} exist, if the spacetime possesses equatorial photon rings. 
The converse result, i.e., inferring from the existence of spherical photon orbits the separability of a space time, 
can be shown to hold only for a limited class of spacetimes satisfying some specific restriction, which we will not discuss 
here (see discussion in \cite{KG_GPprep} for more details on this).\footnote{In short, one cannot exclude the possibility 
of having non-separable spacetimes with spherical photon orbits.}
A corollary of this proposition is that a spacetime \emph{cannot be separable} if it does not admit spherical orbits (in any coordinates) 
while it admits equatorial photon rings.  We now proceed to demonstrate our results. 

For a spacetime of the general form (\ref{metric}) the Hamilton-Jacobi equation for null geodesics becomes~\cite{MTW1973}, 
\be 
\frac{(S_{,r})^2}{g_{rr}}  +\frac{(S_{,\theta})^2}{g_{\theta\theta}}  -V_{\textrm{eff}}=0, 
\label{HJ1}
\ee
where $ S(r,\theta)$ is Hamilton's characteristic function. A separable spacetime entails an additive 
$S$, i.e. $S= S_r(r) + S_\theta (\theta)$~\cite{Landau.mech.book}. Assuming the conditions 
\begin{align}
& g_{\theta\theta} V_{\textrm{eff}}= f_1 (r) h(\theta) + g(\theta), 
\label{sepa_condition1}
\\
\nn \\
& \frac{g_{\theta\theta}}{g_{rr}} = f_2 (r) h(\theta),
\label{sepa_condition2}
\end{align} 
(here $f_1, f_2,h,g$ are arbitrary functions of their argument) we can rearrange (\ref{HJ1}) as, 
\be
f_2(r)  (S_r^\p )^2 - f_1 (r) = \frac{1}{h(\theta)} \left [\, g(\theta) - ( S_{\theta}^\p)^2 \, \right ] = \cC,
\label{HJ2}
\ee
and demonstrate the separability of the system, with $\cC$ playing the role of the third constant (or `Carter constant'). 
The constraints  (\ref{sepa_condition1}), (\ref{sepa_condition2}) are obeyed by Carter's general class of `canonical' 
metrics~\cite{Carter1968CMaPh} that allow the separation of the Hamilton-Jacobi, Schr\"odinger and scalar wave equations. 
Members of this class include known separable metrics such as the Kerr metric and its deformation devised 
by Johannsen~\cite{Johannsen2013PhRvD}.

We can now make contact with spherical photon orbits. For $r_0 = \mbox{const.}$ the
condition (\ref{circularity}) becomes,
\be
\left ( g_{\theta\theta} V_{\rm eff} \right )_{,r} |_{r_0} =0,  
\label{sphericity1}
\ee
which is a condition for the existence of spherical orbits. 
But if a spacetime is separable, then the condition (\ref{sepa_condition1}) holds, which then implies that the condition 
(\ref{sphericity1}) also holds as long as $f_1^{\p}(r_0)=0$ has roots. Therefore a separable spacetime can admit 
spherical photon orbits. If a given spacetime is additionally known to have equatorial photon rings, then this ensures 
that the equation $f_1^{\p}(r)=0$ has roots. 
This can be seen from (\ref{HJ2}), after setting $S_r^\p= g_{rr}u^r$:
\be 
f_2(r)\left(g_{rr}u^r\right)^2=\cC+f_1(r)\equiv V_{r}(r).
\ee
Provided a photon ring exists, then $V_{r}=V_{r}^\p=0$ at the ring's radius. The latter condition implies $f_1^{\p}(r)=0$.
We have thus established the first half of our result.

It is now easy to demonstrate the corollary result: for a separable spacetime with an equatorial photon ring but 
without spherical orbits, the preceding analysis implies the conditions (\ref{sepa_condition1}), (\ref{sepa_condition2}) 
and  $f_1^{\p}(r)=0$ for some $r_0$. In turn this means that Eq. (\ref{sphericity1}) holds and as a consequence $r=r_0$ 
solves the circularity condition, i.e., spherical orbits exist in contradiction with the initial assumption. 

Although the existence of an equatorial photon ring may look like a strong mathematical requirement, 
it is nevertheless a very reasonable physical assumption. Indeed, photon rings emerge as a generic property of black hole solutions 
in GR, known `deformations' of these solutions, UCO spacetimes and so on~\cite{Johannsen2010ApJ,
Cardoso:2016rao,Cunha2017PhysRevLett,Glampedakis2018PhRvD}. 

Another point that should be highlighted is the inherent gauge dependence of spherical orbits. These orbits occur in an 
appropriate coordinate system, like the one that allows separability. In a different set of coordinates these orbits may become 
spheroidal ones with $r_0 = r_0 (\theta)$. It should be made clear therefore that the spherical orbits/separability discussion 
is taking place in specific coordinate system.

A question that goes beyond this section's results concerns the presence of spheroidal orbits in \emph{non-separable} 
axisymmetric-stationary spacetimes. This issue is explored in more detail in a companion paper~\cite{KG_GPprep}; here it will 
suffice to say that it is not easy to ascertain whether such orbits exist or not when separability is lost. The numerical results 
reported in the following section and in~\cite{ KG_GPprep} suggest that spacetimes that are deformed away from separability 
have their photon orbits deformed into spheroidal.


\emph{Loss of spherical orbits in non-Kerr spacetimes.  ---}
In order to explore the implications of our previous results, we consider the specific example
of the non-separable Johannsen-Psaltis (JP) metric, see~\cite{Johannsen:2011dh} for its explicit form
(a more detailed analysis of the JP metric as well as similar results for other metrics can be found in our companion 
paper~\cite{KG_GPprep}). This is a spacetime of the form (\ref{metric}) (albeit not a solution of the 
GR equations) constructed as a deformation away from Kerr. The deformation itself is controlled by a single 
constant parameter $\varepsilon_3$, so that $\varepsilon_3=0$ corresponds to the Kerr metric.

To begin with, it is possible to show rigorously~\cite{KG_GPprep} that the circularity condition~(\ref{circularity}) 
in the JP metric does not admit spherical solutions (in it's standard coordinates).

A hands-on understanding of the degree 
of `decircularisation' suffered by the photon orbits can be gained by performing direct numerical integrations of the 
corresponding geodesic equations. The aim is to test whether photons could be trapped 
in the vicinity of the JP  `black hole'  to anything resembling a spherical orbit.

\begin{figure}[htb!]
\includegraphics[width=0.23\textwidth]{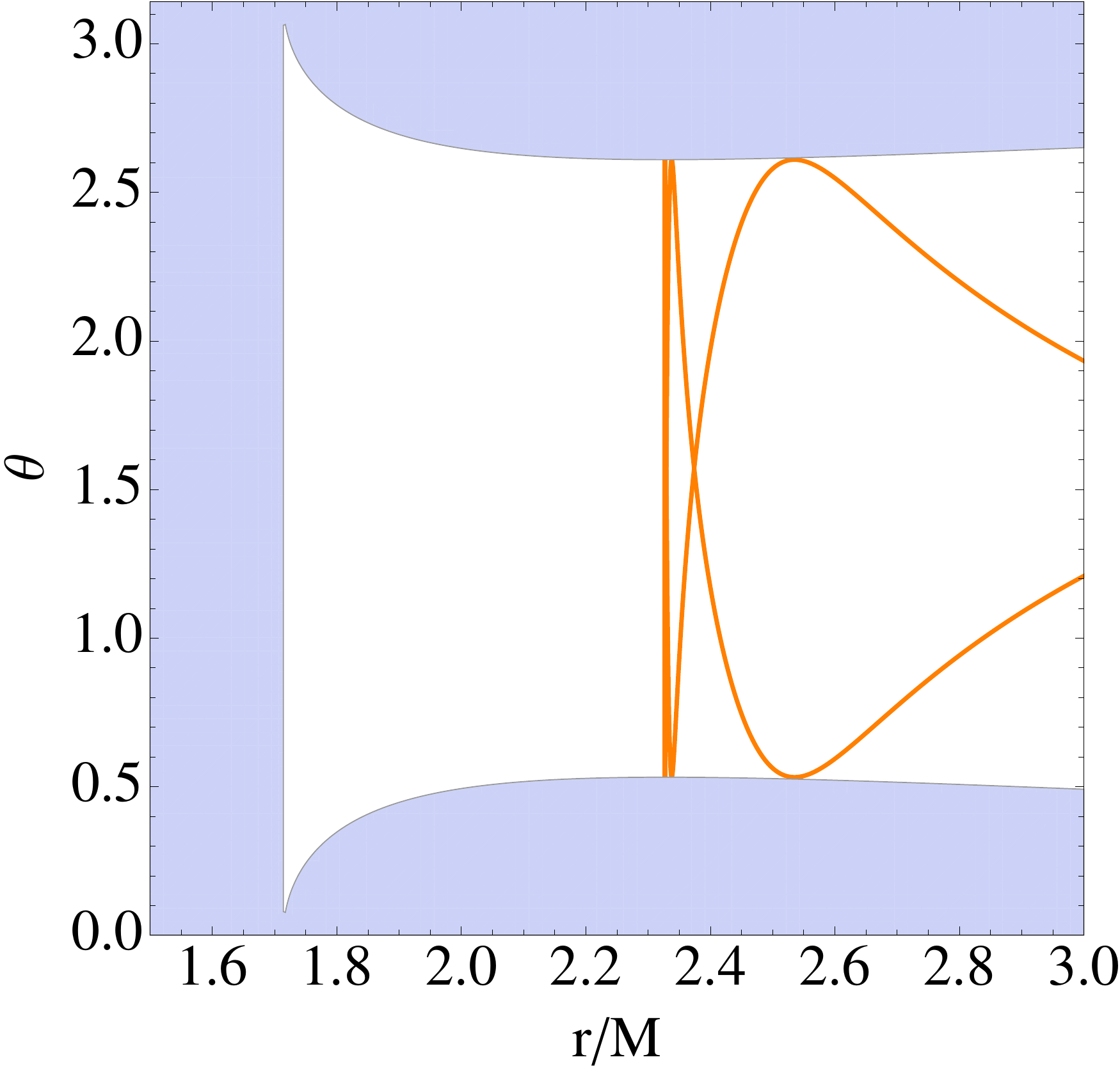} 
\includegraphics[width=0.23\textwidth]{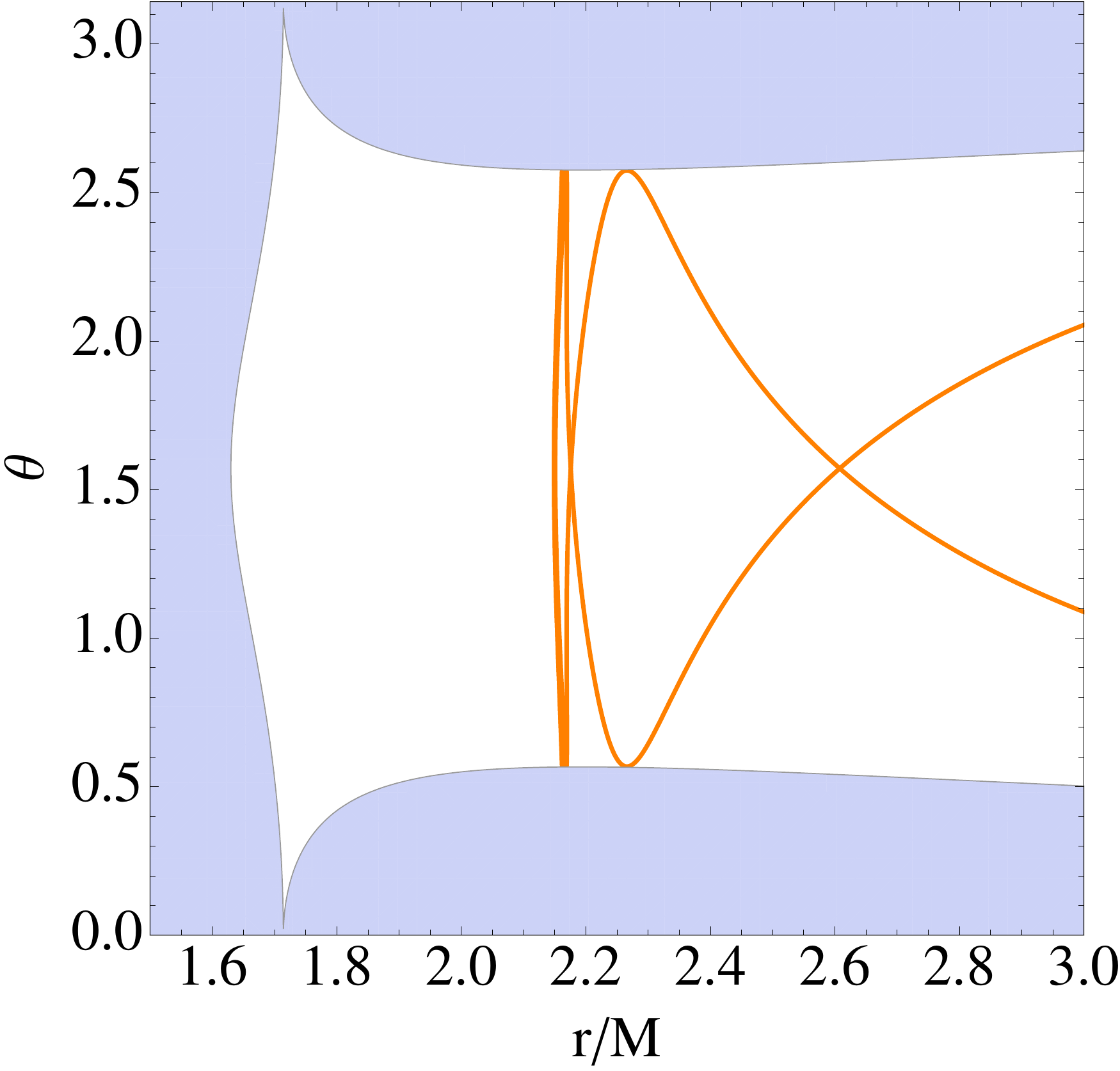} 
\includegraphics[width=0.2365\textwidth]{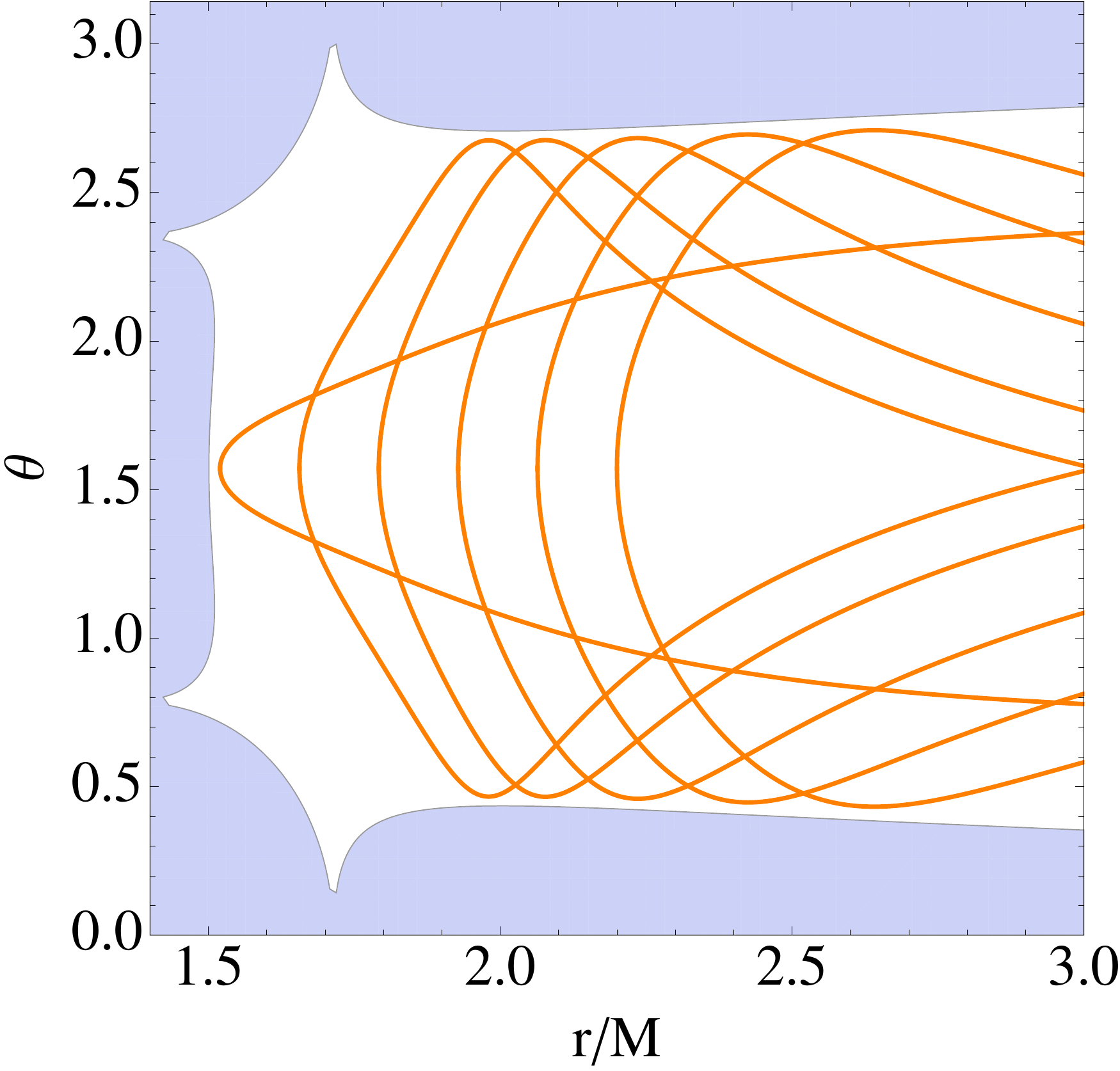}
\includegraphics[width=0.228\textwidth]{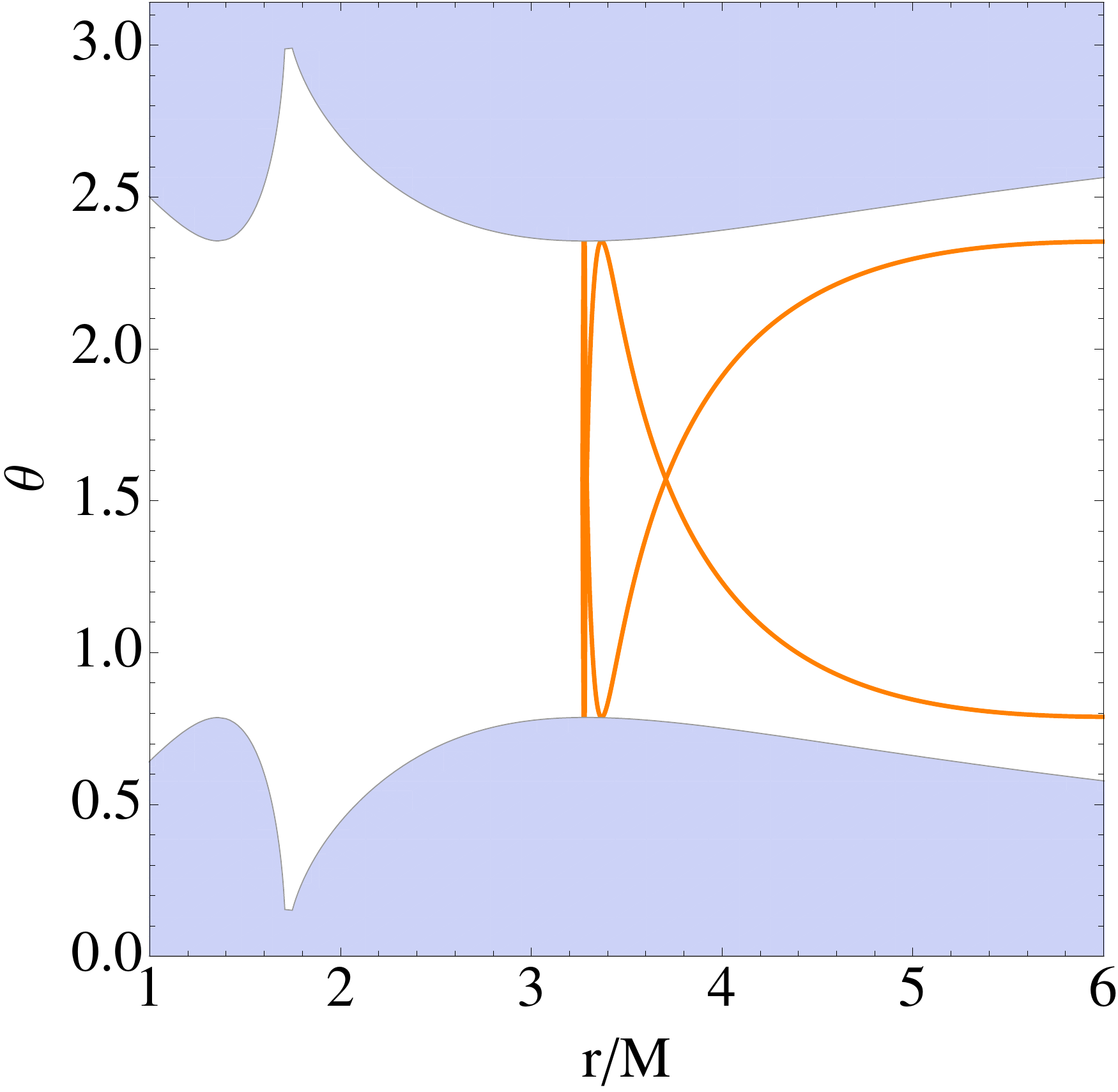}
\caption{\emph{The $r(t), \theta(t)$ profiles of spherical, spheroidal and scattering orbits in the Kerr and JP metric for $a=0.7M$.}
The top row shows co-rotating orbits with impact parameter $b=2.169$ in a Kerr spacetime (left) and in a JP spacetime 
with $\varepsilon_3=1$ (right). The bottom left plot shows plots for several prograde orbits with an impact parameter of $b=1.5$ in a JP 
spacetime, with $\varepsilon_3=5$. The bottom right plot shows a retrograde orbit for a JP spacetime 
again with the same $\varepsilon_3$, but with $b=4.09$. The Kerr orbit remains close to the spherical orbit, 
at about $r_0\approx 2.3268M$.}
\label{fig:r(th)_Kerr}
\end{figure}

\begin{figure}[htb!]
\includegraphics[width=0.4\textwidth]{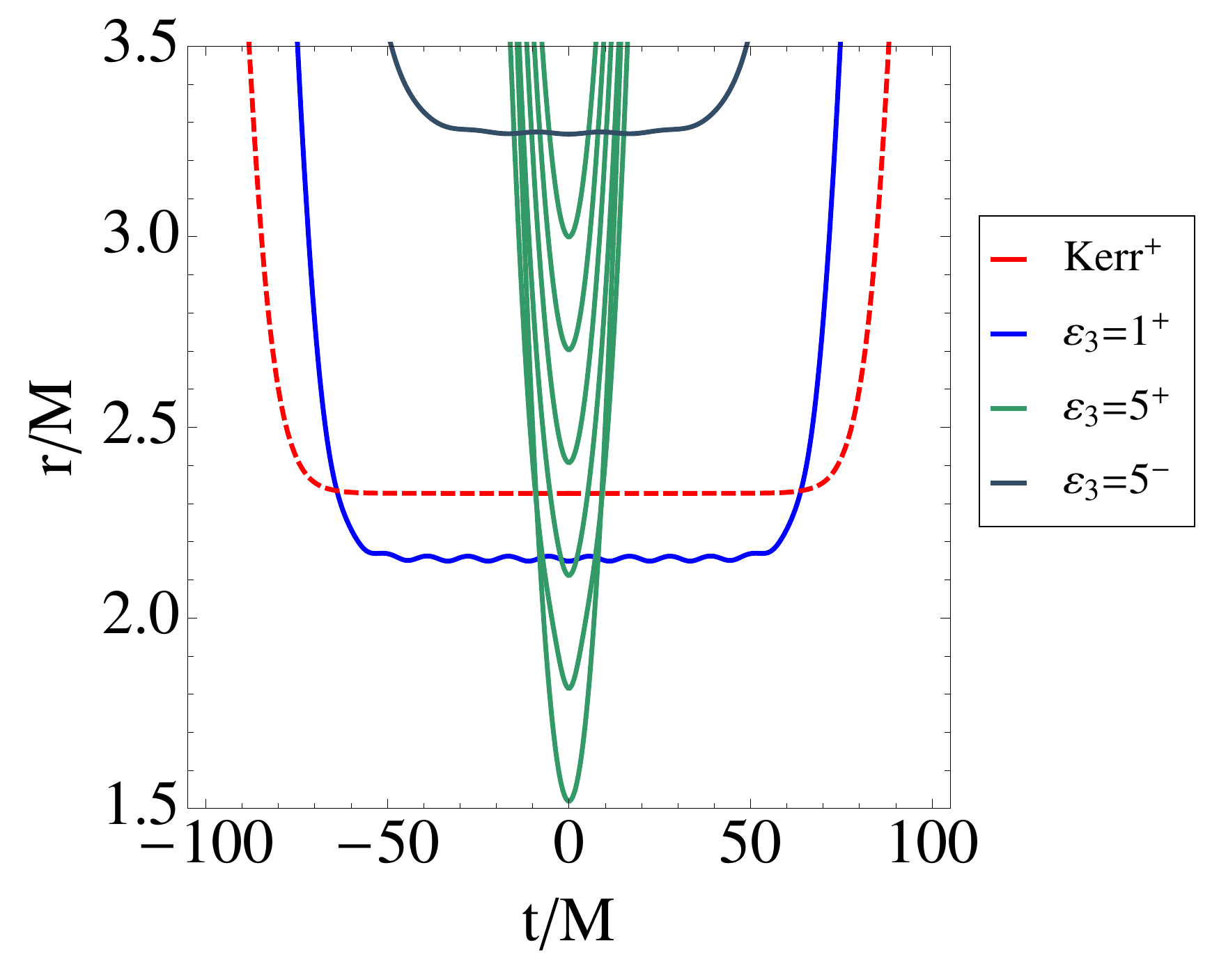}
\caption{\emph{Loss of spherical photon orbits as a function of the deformation away from Kerr.}
We show the $r(t)$ profiles of the JP photon orbits of Fig.~\ref{fig:r(th)_Kerr}, including an example of a spherical
Kerr orbit (red dashed curve). A plus/minus indicates a prograde/retrograde orbit. The prograde $\varepsilon_3=1$ orbit 
(blue wavy curve) as well as the retrograde $\varepsilon=5$ one (dark green curve) are  
spheroidal, resulting in photons 
trapped near the black hole for a considerable period of time. No such notion exists in the case of prograde 
$\varepsilon_3=5$ orbits (green curve).}
\label{fig:r(t)_JP}
\end{figure}

We setup our numerical experiment in the following way. For the particular case of the Kerr metric ($\varepsilon_3=0$) and 
for a given spin parameter, we launch a photon from a point on the equatorial plane, as close as numerically possible to the 
known photon orbit so that it does not plunge into the black hole. We then monitor the photon's trajectory forward and backward 
in time and see how this close-approach `zoom-whirl' orbit looks like. In Kerr, the outcome is the expected one, i.e., the photon 
approaches the spherical photon orbit, remains in its vicinity for a few `whirls' and then moves away from the black hole. 
During the whirling phase, the photon remains with high accuracy on a constant radius. An example of such a Kerr orbit 
(together with the zero-velocity separatrix $V_{\rm eff} = 0$) is show at the top left plot in Fig.~\ref{fig:r(th)_Kerr}.
The orbit's radial profile $r(t)$ is shown in Fig.~\ref{fig:r(t)_JP} (red dashed curve). 

The same procedure is followed in the integration of the $\varepsilon_3 \neq 0$ geodesics, but this time it is done
while we lack the guiding help of a beforehand known photon orbit radius. As soon as an appropriate orbit is found, we try
to push it as close to the plunging limit as possible. 

An example of such a photon orbit for $\varepsilon_3=1$ is shown in Fig.~\ref{fig:r(th)_Kerr} (top right panel).  
At a first glance, the orbit looks Kerr-like. However, a more careful inspection reveals that this is not the case, in agreement 
with our earlier comment about the absence of spherical orbits. The non-Kerr character of the orbit is more visible in 
Fig.~\ref{fig:r(t)_JP}, where $r(t)$ can be seen to take a  
spheroidal form, displaying small oscillations about some mean radius. 
This result offers a nice example of a spacetime that has no spherical orbits  
but can, nevertheless, trap photons in  
spheroidal orbits for considerable time intervals. 

The situation can be markedly different for a larger deformation, e.g. $\varepsilon_3=5$. Any attempt to find 
a  
spheroidal orbit in this JP metric results in failure, provided the orbits are \emph{prograde} ($b>0$) 
and of low inclination with respect to the equatorial plane. 
Some examples of failed attempts can be seen in Fig.~\ref{fig:r(th)_Kerr} (bottom left panel). A typical $r(t)$ profile
can be seen in Fig.~(\ref{fig:r(t)_JP}); in the same figure it is also evident the inability of the photon to spend any 
appreciable time near the black hole.
In contrast, \emph{retrograde} orbits ($b<0$) appear to behave in the  
spheroidal fashion of the previous $\varepsilon_3=1$ 
orbits, see bottom right panel of Figure~\ref{fig:r(th)_Kerr} and Fig.~(\ref{fig:r(t)_JP}). This is not entirely surprising, since a photon 
in a retrograde orbit would stay further out and be less affected by the spacetime's non-Kerr deformations.  
%

The upshot of this discussion is that although non-separability may result in the loss of spherical orbits,  
photons could still be temporarily trapped in  
spheroidal trajectories if the departure from separability is in some sense ``small". 
Furthermore, significantly deformed non-Kerr spacetimes may display a much richer phenomenology such as 
non-equatorial photon orbits (see \cite{KG_GPprep} for details).


\emph{Implications for non-Kerr objects.---} 
The topic of this letter, namely, the connection between separability of a spacetime and the behaviour of photon 
orbits (spherical, spheroidal or non-trapping), can become a powerful 
(and to some extent a model/theory-agnostic) tool for diagnosing the presence of a non-Kerr metric.

Black hole ringdown signals (as the ones produced by merging black holes) are dominated by the $(\ell,m)=(2,2)$ 
QNM, associated with the (prograde) equatorial photon ring. However, tests of the GR no-hair theorem requires the simultaneous 
observation of additional modes, with $(3,3), (2,1), (3,2)$ being the most likely candidates, in terms of detectability~\cite{Berti:2016lat,Baibhav2018}. 
Our results have a direct impact on the $0< m <\ell$ modes which are naturally associated with prograde spherical photon orbits. 
Assuming that the merger produces a non-Kerr object with a non-separable spacetime, its QNM spectroscopy might 
reveal a number of  missing/dimmed `lines' that would otherwise manifest as $\ell\geq m$ Kerr QNMs. This could be smoking gun
evidence for a non-Kerr object. 

In parallel with GW observations, one could look for the electromagnetic signature of absent spherical photon orbits in the 
shadow formed by supermassive black holes when illuminated by their radiating accretion flows. The goal of the EHT project 
is to produce a sharp image of the shadow of our galactic Sgr A$^*$ black hole candidate (and of other nearby systems)~\cite{Broderick2014ApJ}. 
The morphology of the shadow largely reflects the existence of photon rings and spherical photon orbits in the underlying spacetime. 
This detailed imaging should be able to identify the presence of spherical photon orbits or perhaps the lack thereof 
(in the form of missing trapping orbits). 

The implications of this work are clearly intriguing and potentially important, and should be addressed by future work.  
As a first extension, in a forthcoming paper we plan to investigate the scattering of scalar waves in a non-Kerr spacetime 
with the purpose of understanding how the wave dynamics is affected by the loss of separability and spherical orbits.


\emph{Acknowledgments}.\,We thank Theocharis Apostolatos and Georgios Lukes-Gerakopoulos for valuable 
feedback, and acknowledge support from the COST Actions GWverse CA16104 and PHAROS  CA16214. 
GP acknowledges financial support provided under the European Union's H2020 ERC, Starting Grant agreement 
no.~DarkGRA--757480.


\bibliography{biblio.bib}


\end{document}